\documentclass[11pt]{article}
\pdfoutput=0

\usepackage{epsfig}
\usepackage{amssymb}
\usepackage{graphics,graphpap}

\usepackage{bbm}
\usepackage{amsmath}
\usepackage[a4paper]{hyperref}

\usepackage{color}

\renewcommand{\thefootnote}{\fnsymbol{footnote}}

\newcommand\T{\rule{0pt}{2.5 ex}}
\newcommand\Bo{\rule[-1.5ex]{0pt}{0pt}}

\def \sla#1{#1\!\!\!/}

\begin{document}

\begin{titlepage}

\vskip3.cm

\begin{center}
{\Large \bf \boldmath
Two-loop Corrections to the $B\to\pi$ Form Factor from QCD Sum Rules on the 
Light-Cone and $|V_{ub}|$
}

\vskip2cm 

{\sc
Aoife Bharucha\footnote{aoife.bharucha@desy.de}
} \vskip0.5cm
{\em 
 II. Institut f\"ur Theoretische Physik,
 Universit\"at Hamburg,
Luruper Chaussee 149,
22761 Hamburg, Germany}

\vskip3cm

{\large\bf Abstract\\[10pt]} \parbox[t]{\textwidth}{
We calculate the leading-twist $\mathcal{O}(\alpha_s^2 \beta_0)$ corrections to the $B\to\pi$ transition form factor $f_+(0)$ in light-cone sum rules. We find that, as expected, there is a cancellation between the $\mathcal{O}(\alpha_s^2 \beta_0)$ corrections to $f_B f_+(0)$ and the large corresponding corrections to $f_B$, calculated in QCD sum rules. This suggests the insensitivity of the form factors calculated in the light-cone sum rules approach to this source of radiative corrections. We further obtain an improved determination of the CKM matrix element $|V_{ub}|$, using latest results from BaBar and Belle for $f_+(0)|V_{ub}|$.
}

\end{center}
\end{titlepage}

\setcounter{footnote}{0}
\renewcommand{\thefootnote}{\arabic{footnote}}
\renewcommand{\theequation}{\arabic{equation}}

\newpage

\section{Introduction}\label{sec:1}

In the last decade we have witnessed major advances in the efforts to overconstrain the sides of the unitarity triangle, in order to test the CKM (Cabibbo-Kobayashi-Maskawa) mechanism of the Standard Model (SM). However, one side of the common parameterisation of this triangle is given by $|V_{ub}|/|V_{cb}|$, where $V_{ij}$ are elements of the CKM matrix, and recent determinations of $|V_{ub}|$ have uncertainties of approximately 10\%~\cite{Babar:2010uj}, as opposed to the error on measurements of $|V_{cb}|$ from the inclusive channel $B\to X_c l \nu$ which is below 2\%~\cite{Barberio:2008fa}. Since the inclusive channel $b\to u l \nu$ is dominated by the large $b\to c l \nu$ background, a competitive determination of $|V_{ub}|$, promising both theoretically and experimentally, is found via the exclusive semi-leptonic decay $B\to\pi l\nu$. This requires information about the relevant hadronic matrix element, parameterised by the form factors $f_+(q^2)$ and $f_-(q^2)$,
\begin{equation} 
\langle\pi(p)|\bar{u}\gamma_\mu b|B(p_B)\rangle=(p_B+p)_\mu f_+(q^2)+(p_B-p)_\mu f_-(q^2),
\end{equation}
where $p_B$ and $p$ are the momenta of the $B$ and $\pi$ mesons respectively and $q^2=(p_B-p)^2$. The beauty of this channel lies in the fact that in the limit of massless leptons, applicable to $l=e$ and $\mu$, only $f_+(q^2)$ is required~\cite{Ball:2005tb},
\begin{equation}
 \frac{d\Gamma}{dq^2}(B^0\to\pi^-l^+\nu_l)=\frac{G_F^2 |V_{ub}|^2}{192 \pi^3 m_B^3} \lambda^{3/2}(q^2) |f_+(q^2)|^2,
\end{equation} 
where $G_F$ is the Fermi coupling constant and $\lambda(q^2)=(m_B^2+m_\pi^2-q^2)^2-4 m_B^2 m_\pi^2$ for masses $m_B$ and $m_\pi$ of the $B$ and $\pi$ mesons respectively. Therefore the extraction of $|V_{ub}|$ relies on the theoretical prediction for a single hadronic quantity $f_+(q^2)$, possible via non-perturbative techniques such as Lattice quantum chromodynamics (QCD) (see e.g. Refs.~\cite{Dalgic:2006dt,Bailey:2008wp}) or QCD sum rules on the light-cone (LCSR).

Theoretical predictions are usually confined to a particular region of $q^2$, for example LCSR are restricted to large recoil energies of the pion, corresponding to $q^2\lesssim 6-7\,\mathrm{GeV}^2$, and Lattice results to small values of the pion momentum\footnote{Note that the form factor at $q^2=0\,\mathrm{GeV}^2$ was recently obtained in a quenched calculation on a very fine lattice~\cite{AlHaydari:2009zr}.}, i.e. $q^2\gtrsim 15\,\mathrm{GeV}^2$. Experimentally the $q^2$ distribution has been measured with increasing accuracy at CLEO~\cite{Athar:2003yg,Adam:2007pv}, BaBar~\cite{Aubert:2005cd,Aubert:2006px,Babar:2008gka,delAmoSanchez:2010zd,Babar:2010uj} and Belle~\cite{Belle:2008kn,Ha:2010rf}. In order to maximally exploit these theoretical and experimental results, one requires a well motivated parameterisation for the $q^2$ dependence of $f_+(q^2)$. There are a number of approaches, either simple pole-type parameterisations as in Refs.~\cite{Becirevic:1999kt,BZ04}, using dispersive bounds to constrain the coefficients of a series expansion as in Refs.~\cite{Boyd:1994tt,Bourrely:2008za} or using the Omn\`{e}s representation as in Refs.~\cite{Flynn:2007qd,Flynn:2007ii}. In all these, the normalisation provided by the LCSR prediction at $q^2=0\,\mathrm{GeV}^2$ plays a crucial role. In fact, one can obtain $|V_{ub}|$ directly from the model independent result for $|V_{ub}|f_+(0)$, first calculated in Ref.~\cite{Ball:2006jz} by fitting such shape parameterisations to BaBar data~\cite{Aubert:2006px}.

Light-cone sum rules are an adaptation of the traditional QCD sum rules approach~\cite{Shifman:1978bx,Shifman:1978by}, considering instead the correlator of the T product of two quark currents sandwiched between a final on-shell meson and the vacuum~\cite{Balitsky:1989ry,Chernyak:1990ag}. This can be expanded about the light-cone, in terms of perturbatively calculable hard scattering kernels convoluted with non-perturbative, universal light-cone distribution amplitudes. The correlator can also be expressed as the sum over excited states, the first being the $B$ meson which is then followed by a continuum of states. Then assuming quark hadron duality above a certain continuum threshold, one can subtract this continuum contribution from both sides. Borel transforming this relation then ensures that this assumption, and the truncation of the series, have a minimal effect on the resulting sum rule.

We are interested in calculating the subset of two-loop radiative corrections to $f_+(0)$ proportional to $\beta_0$, assuming, as discussed in Sec.~\ref{sec:4}, that this is a good approximation to the complete next-to-next-to-leading order (NNLO) result. In addition to allowing an improved determination of $|V_{ub}|$, our calculation will enable us to investigate the size of these radiative corrections in view of the sizeable two-loop contribution to $f_B$ in QCD sum rules~\cite{Jamin:2001fw,Penin:2001ux}. The magnitude of this contribution is thought to be due to coulombic corrections, as explored in e.g. Ref.~\cite{Bagan:1991sg}. The LCSR approach to form factors involves taking the ratio of $f_B f_+(q^2)$, also affected by such coulombic corrections, to $f_B$. We therefore test the argument that radiative corrections should cancel in this ratio, provided both quantities are calculated in sum rules.

The current status of the LCSR calculation of $f_+(q^2)$ is as follows. The next-to-leading order (NLO) twist-2 corrections to $f_+(q^2)$ were first calculated in LCSR in Ref.~\cite{BBB98} and the leading order (LO) corrections up to twist-4 were calculated in Ref.~\cite{Khodjamirian:2000ds}. Since the LO twist-3 contribution was found to be large, further improvements were made by calculating the smaller NLO corrections~\cite{BZ04}. A more recent update where the $\overline{\rm MS}$ mass is used in place of the pole mass for $m_b$ can be found in Ref.~\cite{Duplancic:2008zz,Khodjamirian:2011ub}.

The following paper is structured as follows: in Sec.~\ref{sec:2} we introduce the necessary notation and establish the framework required for the calculation, including the expression for the one-loop correction at leading-twist; in Sec.~\ref{sec:3} we present details of the two-loop calculation and describe the structure of the divergences of the bare result and the renormalisation procedure; a detailed analysis of our numerical results, with predictions for $|V_{ub}|$, can be found in Sec.~\ref{sec:6}; finally we summarise in Sec.~\ref{sec:7}.

\section{Set-up of the calculation}\label{sec:2}
Such as to briefly introduce the LCSR approach to the calculation of $f_+(q^2)$, and the notation which will later be required, we consider the correlator of two quark currents sandwiched between the vacuum and pion,
\begin{align}
\label{eq:correlator} \Pi_\mu&=i\,m_b\int d^D x e^{-i\,p_B\cdot x} \langle\pi(p)| T \{\bar{u}(0)\gamma_\mu b(0)\bar{b}(x)i \gamma_5d(x)\}|0\rangle,\\
\label{eq:correlator2}&=(p_B+p)_\mu \Pi_+(p_B^2,q^2)+(p_B-p)_\mu \Pi_-(p_B^2,q^2).
\end{align}
In the region around the pole at $p_B^2=m_B^2$, $\Pi_+(p_B^2,q^2)$ can be expressed in terms of $f_+(q^2)$ and the $B$ meson decay constant $f_B$, where
\begin{equation}
m_b\langle 0|\bar{d}i\gamma_5 b|B\rangle=m_B^2 f_B.
\end{equation}
Above the $B$ meson pole the contribution of the hadronic states can be described by the spectral density $\rho_{\rm had}$, leading to an expression for the correlator of the form
\begin{equation}
\label{eq:PiHad}
\Pi_+(p_B^2,q^2)=f_B m_B^2 \frac{f_+(q^2)}{m_B^2-p_B^2}+\int_{s>m_B^2} ds \frac{\rho_{\rm had}}{s-p_B^2}.
\end{equation}
Alternatively, in the Euclidean region where $p_B^2-m_B^2$ is large and negative, using a light-cone expansion about $x^2=0$, the correlator can be collinearly factorised into perturbatively calculable hard kernels $\mathcal{T}_+^{(n)}(u,\mu^2)$ and non-perturbative light-cone distribution amplitudes (DAs) $\phi^{(n)}(u,\mu^2)$ for a given twist $n$, via
\begin{equation}\label{eq:conv}
\Pi_+(p_B^2,q^2)=\sum_n \int du\, \mathcal{T_+}^{(n)}(u,p_B^2,q^2,\mu^2)\phi^{(n)}(u,\mu^2),
\end{equation}  
where $u$ is the momentum fraction of the quark in the pion, and $\mu$ is the factorisation or renormalisation scale. This factorisation theorem is not proved to all orders, but can be verified at a given order in twist or perturbation theory by the cancellation of IR and soft divergences, the latter arising when the convolution does not converge at the endpoints.
The leading-twist pion distribution amplitude, $\phi(u,\mu^2)$, contains the distribution of the momentum fraction $u$ in the pion's infinite momentum frame for the lowest Fock state. We postpone the discussion of DAs to Sec. \ref{sec:LCDAs}, and here simply state the definition, in the Fock-Schwinger or light-cone gauge, to be 
\begin{equation} 
\label{eq:phi}
\langle \pi(p)|\bar{u}(0)\gamma_\mu\gamma_5\,d(x)|0\rangle=-i f_\pi p_\mu\int_0^1du\,e^{i \bar{u} p\cdot x}\phi(u,\mu^2)+\dots,
\end{equation} 
where $f_\pi$ is the decay constant of the pion, $\bar{u}=1-u$ is the momentum fraction of the antiquark, and the ellipsis indicates the contributions at higher-twist.
Making the substitution $u=(m_b^2-q^2)/(s-q^2)$ in the leading twist contribution to Eq.~(\ref{eq:conv}), and taking the imaginary part, we can define the spectral density $\rho_{\mathrm{T2}}$ at twist-2,
\begin{equation}
\label{eq:PiLC}
\Pi_+(p_B^2,q^2)=\int_0^\infty ds \frac{\rho_{\mathrm{T2}}}{s-p_B^2}+\dots,
\end{equation}
where again the ellipsis indicates the contributions at higher-twist. Equating the expressions for $\Pi_+(p_B^2,q^2)$ in Eqs.~(\ref{eq:PiHad}) and (\ref{eq:PiLC}) results in 
\begin{equation}
f_B m_B^2 \frac{f_+(q^2)}{m_B^2-p_B^2}+\int_{s>m_B^2} ds \frac{\rho_{\rm had}}{s-p_B^2}=\int_0^\infty ds \frac{\rho_{\mathrm{T2}}}{s-p_B^2}.
\label{eq:LCSRcont}
\end{equation}
Above the continuum threshold $s_0$, a continuum of states contributes and the approximation of quark-hadron duality is thought to be reasonable, such that
\begin{equation}
 \rho_{\rm had}=\rho_{\mathrm{T2}}\,\Theta(s-s_0).
\end{equation}
Subtracting the continuum contribution and Borel transforming both sides results in the sum rule for $f_+(q^2)$,
\begin{equation}
 f_+(q^2)=\frac{1}{f_B m_B^2}\int_{m_b^2}^{s_0} ds\, \rho_{\mathrm{T2}}\,e^{-(s-m_B^2)/M^2},
\end{equation}
where $M^2$ is the Borel parameter. The uncertainty introduced in making the quark-hadron duality approximation is reduced by Borel transforming, and further by choosing $s_0$ and $M^2$ appropriately such that the result for  $f_+(q^2)$ is flat with respect to these parameters.

Returning to the original definition of the correlator in Eq.~(\ref{eq:correlator}), we consider the NLO corrections to the leading-twist term in the expansion about the light-cone $x^2=0$, calculated in Ref.~\cite{BBB98}. In analogy to Eq.~(\ref{eq:conv}), we express the correlator in the collinearly factorised form,
\begin{equation}\label{eq:convmu}
\Pi_\mu(p_B^2,q^2)=\sum_n \int du\, \mathcal{T}_\mu^{(n)}(u,\mu^2)\phi^{(n)}(u,\mu^2).
\end{equation} 
We perturbatively expand the leading-twist contribution to the correlator, 
\begin{align}
\Pi^{\mathrm{T2}}_\mu&=\int du\, \mathcal{T}_\mu^{(2)}(u,\mu^2)\phi(u,\mu^2)\\
 &=\Pi_\mu^{(0)}+\frac{\alpha_s}{4 \pi}\,\Pi_\mu^{(1)}+\left(\frac{\alpha_s}{4 \pi}\right)^2 N_f\,\Pi_\mu^{(2)}\dots\,,
\label{eq:PiT2}
\end{align}
where the tree-level term $\Pi_\mu^{(0)}$ is
\begin{equation}
 \Pi_\mu^{(0)}=-\frac{1}{4}\,f_\pi\,m_b\int_0^1du\,\phi(u,\mu^2)\,tr\{\gamma_\mu \frac{\not{p}_B-\bar{u}\sla{p}+m_b}{(p_B-\bar{u} p)^2-m_b^2} \sla{p}\}.
\end{equation}
Although the $\mathcal{O}(\alpha_s)$ radiative corrections to the correlator, involving six further diagrams, were calculated in Ref.~\cite{BBB98}, we include the following expressions here as they will be useful in presenting the NNLO results,
\begin{equation}
\Pi_\mu^{(1)}=\frac{\mathcal{N}}{4} \int_0^1du\,\phi(u,\mu^2)\int\frac{d^D k}{(2\pi)^D} \frac{g^{\alpha\beta}}{k^2} F^{\mathrm{T}}_\mu,
\end{equation}
where the normalisation $\mathcal{N}$ is defined as
\begin{equation}
\mathcal{N}=-i\,(4 \pi)^2\,C_F f_\pi\,m_b,
\end{equation}
for $C_F=4/3$. $F^{\mathrm{T}}_\mu$ contains the total contribution of the traces and fermionic propagators for the weak vertex correction, $B$ vertex correction, box, $b$ quark self-energy and light quark self-energy diagrams. We factorise the gluon propagator out of $F^{\mathrm{T}}_\mu$ so that our notation can be adapted to the NNLO calculation more easily. Defining $F^{\mathrm{T}}_\mu$ to be
\begin{equation}\label{eq:FT} 
F^{\mathrm{T}}_\mu=F^{\mathrm{WV}}_\mu+F^{\mathrm{BV}}_\mu+F^{\mathrm{BX}}_\mu+F^{\mathrm{SE}}_\mu+F^{\mathrm{LSE}}_\mu,
\end{equation}
the contribution of individual diagrams in Feynman gauge can be expressed as
\begin{eqnarray}
F^{\mathrm{WV}}_\mu&=&tr\{\gamma_\alpha\frac{\sla{k}-u\sla{p}}{(k-up)^2}\gamma_\mu\frac{\sla{q}-\sla{k}+u\sla{p}+m_b}{ (q-k+up)^2-m_b^2}\gamma_\beta \frac{\sla{p}_B-\bar{u}\sla{p}+m_b}{(p_B-\bar{u} p)^2-m_b^2} \sla{p}\}\\
F^{\mathrm{BV}}_\mu&=&tr\{\gamma_\mu \frac{\sla{p}_B-\bar{u}\sla{p}+m_b}{(p_B-\bar{u} p)^2-m_b^2} \gamma_\alpha\frac{-\sla{p}_B-\sla{k}+\bar{u}\sla{p}-m_b}{ (p_B+k-\bar{u}p)^2-m_b^2}\frac{\sla{k}-\bar{u}\sla{p}}{(k-\bar{u}p)^2}\gamma_\beta\,\sla{p}\}\\
F^{\mathrm{BX}}_\mu&=&tr\{\gamma_\alpha \frac{u\sla{p}-\sla{k}}{(up-k)^2}\gamma_\mu \frac{\sla{p}_B-\bar{u}\sla{p}-\sla{k}+m_b}{(p_B-\bar{u} p-k)^2-m_b^2}\frac{\sla{k}+\bar{u}\sla{p}}{(k+\bar{u}p)^2}\gamma_\beta\, \sla{p}\}\\
F^{\mathrm{SE}}_\mu&=&tr\{\gamma_\mu \frac{\sla{p}_B-\bar{u}\sla{p}+m_b}{(p_B-\bar{u} p)^2-m_b^2} \gamma_\alpha\frac{-\sla{p}_B+\bar{u}\sla{p}+\sla{k}-m_b}{(p_B-\bar{u} p-k)^2-m_b^2} \gamma_\beta \frac{\sla{p}_B-\bar{u}\sla{p}+m_b}{(p_B-\bar{u} p)^2-m_b^2}  \sla{p}\}.
\end{eqnarray}
As in previous calculations, we work in the limit that the light quarks are massless, i.e. $p^2=0$. Therefore $F^{\mathrm{LSE}}_\mu$, the  contribution of the self-energy diagrams for the external light quarks, vanishes as discussed in Sec.~\ref{sec:5}. In this paper, to avoid repeating what already exists in the literature, we will only concentrate on the technical details for the $\mathcal{O}(\alpha_s^2\beta_0)$ corrections. Details of the NLO and higher twist contributions incorporated into our numerical analysis are as given explicitly in Ref.~\cite{BZ04}.

\section{Radiative corrections at order $\alpha_s^2\beta_0$}\label{sec:3}
\subsection{Calculation of the fermion bubble diagrams}\label{sec:4}
In analogy to QED, where the running of the $\beta$-function is connected to the photon polarisation, Brodsky, Lepage and Mackenzie had the idea of associating the running of the QCD $\beta$-function with fermion loop insertions in the lowest order corrections~\cite{Brodsky:1982gc}. The scale for a given process can then be set by demanding that this contribution to the two-loop corrections vanishes, a procedure known as BLM scale setting. Physically, such a renormalisation scale reflects the mean virtuality of the gluon propagator~\cite{Brodsky:1997dh}.

\begin{figure}[t]
\begin{center}
\includegraphics[width=.86\textwidth]{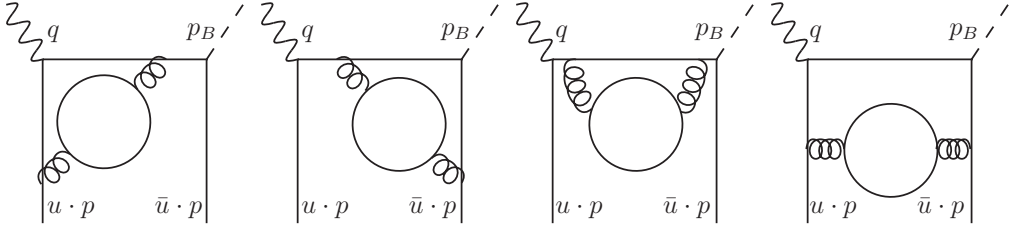}
\end{center}
\caption{Feynman diagrams for $\mathcal{O}(\alpha_s^2\beta_0)$ corrections to  $\Pi_\mu^{T2}$. From left to right, the $B$ vertex correction, weak vertex correction, box and $b$ quark self-energy diagrams are shown. The external quarks are on-shell with momenta as indicated and the dashed line represents the $B$ meson.}\label{fig:diags}
\end{figure}

In Ref.~\cite{Broadhurst:1994se}, the technique of na\"{i}ve non-abelianisation (NNA) was proposed, where the complete NNLO result is approximated by calculating fermion loop insertions, as for BLM scale setting, and replacing $N_f$ by its non-abelian counterpart $-(3/2)\beta_0$. This idea was supported by the observation that in a number of cases where the remaining part of the two-loop corrections could be calculated e.g. higher order corrections to observables from hadronic vacuum polarisation and to the pole mass, it was found to be small in comparison to the $\mathcal{O}(\alpha^s\beta_0)$ contribution\footnote{Further, in Refs.~\cite{Beneke:1994qe,BBB95}, this idea was used to extend the BLM scale setting, by resumming fermion loop insertions in the lowest order corrections to all orders.}. Using the NNA technique, we calculate the $\mathcal{O}(\alpha_s^2\beta_0)$ twist-2 contribution to $f_+(0)$, keeping in mind that the NLO corrections to the higher twist contributions have been found to be comparatively small\footnote{Note that the various contributions to $f^+(0)$ were studied in Ref.~\cite{Duplancic:2008zz} in the pole and $\overline{\mathrm MS}$ schemes, and while at LO the twist-3 are comparable to the LO twist-2 contributions ($\sim 40-50\%$), at NLO, in comparison to the twist-2 ($\sim 10-20\%$), the twist-3 contributions are better under control ($\sim 2-4\%$).}. The expression to be calculated takes the form,
\begin{equation}
 \Pi_\mu^{(2)}=\mathcal{N}\int_0^1du\,\phi(u,\mu^2)\int\frac{d^D k}{(2\pi)^D} \frac{ \Gamma(\epsilon)\Gamma(2-\epsilon)^2}{\Gamma(4-2\epsilon)} \left(\frac{-k^2}{4\pi \mu^2}\right)^{-\epsilon}\frac{1}{k^2}\left(g^{\alpha\beta}-\frac{k^\alpha k^\beta}{k^2}\right)F^{\mathrm{T}}_\mu,
\end{equation}
where $F^{\mathrm{T}}_\mu$ is as defined in Eq.~(\ref{eq:FT}). The relevant Feynman diagrams are shown in Fig.~\ref{fig:diags}. 

The calculation is similar to the one-loop case, however, the additional fermion loop induces two important changes. Firstly, the tensor structure of the gluon propagator changes from the form
\begin{equation}
\frac{-ig^{\alpha\beta}}{k^2}\to\frac{-i}{k^2}\left(g^{\alpha\beta}-\frac{k^\alpha k^\beta}{k^2}\right)
\end{equation}
resulting in additional terms in the trace (although these cancel in the sum of all diagrams due to gauge invariance~\cite{Melic:2001wb}, serving as an additional check of the calculation). Secondly, the factor $\Gamma(\epsilon)$ means that the integrals must be expanded to a higher order in $\epsilon$. The increased complexity of the calculation is slightly compensated by the fact that we set $q^2=0$, however two scales ($p_B^2$ and $m_b$) and one dimensionless parameter ($u$) remain. We perform the traces using the package \texttt{FeynCalc}\cite{Mertig:1990an}, and expand the hypergeometric functions using the Mathematica package \texttt{HypExp}~\cite{Huber:2007dx}. The resulting analytic expression must then be simplified and rearranged into a form facilitating the convolution with the distribution amplitude.

\subsection{Structure of the divergences}\label{sec:5}

The  bare $\mathcal{O}(\alpha_s^2 N_f)$ results for $\Pi^{(2)}_\mu$, contain both infra-red (IR) and ultra-violet (UV) divergences. These are treated in na\"{i}ve dimensional regularisation (NDR), with totally anti-commuting $\gamma_5$ due to the presence of two $\gamma_5$ matrices in the trace, renormalising the UV divergences in the $\overline{\rm MS}$ scheme. As mentioned earlier, in NDR the light quark self energy diagrams vanish, as the UV and IR divergences arising from these diagrams cancel.
On adding all the diagrams together, we first perform the gluon self-energy renormalisation using the $\mathcal{O}(\alpha_s N_f)$ contribution, $Z_{3\mathrm{YM}}^{(1)}$, to the corresponding  renormalisation constant $ Z_{3\mathrm{YM}}$~\cite{Pascual:1984zb},
\begin{equation}
 Z_{3\mathrm{YM}}^{(1)}=-C_F\left(\frac{2}{3\epsilon}\right),
\end{equation}
 multiplied by $\Pi_\mu^{(1)}$. 
The left-over UV poles are completely removed by mass renormalisation, using the $\mathcal{O}(\alpha_s^2 N_f)$ contribution, $Z_m^{(2)}$, to the renormalisation constants $Z_m$,
\begin{equation}
Z_m^{(2)}=C_F \left(-\frac{1}{\epsilon}+\frac{5}{6 \epsilon}\right),
\end{equation}
multiplied by $\Pi_\mu^{(0)}$. Collecting what we assume to be the remaining IR divergences in $\Pi_\mu^{(2),T_{\mathrm{IR}}}$ and subtracting this quantity,
\begin{equation}
\Pi_\mu^{(2),\mathrm{ren.}}=\Pi_\mu^{(2)}- Z_{3\mathrm{YM}}^{(1)}\Pi_\mu^{(1)}-Z_m^{(2)}\Pi_\mu^{(0)}-\Pi_\mu^{(2),T_{\mathrm{IR}}},
\end{equation}
leaves $\Pi_\mu^{(2),\mathrm{ren.}}$ UV and IR finite, however we are still to determine the origin of the IR divergences contained in $\Pi_\mu^{(2),T_{\mathrm{IR}}}$.

\subsection{Convolution and scale dependence}\label{sec:LCDAs}
The leading-twist pion DA defined in Eq.~(\ref{eq:phi}) can be expanded in a series of Gegenbauer polynomials,
\begin{equation}
\label{eq:DAexp}
\phi(u,\mu^2)=6u(1-u)\sum_{n=0}^\infty a_n(\mu^2) C_n^{3/2}(2u-1).
\end{equation}
Here $a_n$ are known as Gegenbauer moments, and in the case of the pion the odd moments are zero by G-parity. The expansion is usually truncated, as the higher moments are suppressed due to the highly oscillatory behaviour of the Gegenbauer polynomials. However, the truncation is only justified if the hard scattering kernel $\mathcal{T}^{(n)}_\mu$ is slowly varying and non-singular for all $u$~\cite{Ball:2005ei}. We include terms for $n\leq 4$ up to $\mathcal{O}(\alpha_s)$, but we assume that at $\mathcal{O}(\alpha_s^2N_f)$ the effect of $a_{2,4}(\mu)$ is negligible\footnote{This can be inferred from Fig. 1 of Ref.~\cite{Ball:2005tb}, where the respective size of different contributions in $a_n$ to $f_+(q^2)$ were shown as a function of $q^2$.}, and adopt the asymptotic DA (i.e. $\phi(u,\infty)=6u(1-u)$) to simplify the convolution.

As the previously calculated twist-3 and 4 contributions are included in our numerical analysis, the corresponding DAs are also required, as defined in Ref.~\cite{BZ04}. In the same reference it was shown that, for a given twist, the two and three particle distribution amplitudes can be related by an equation of motion, resulting in a reduced number of independent parameters: $\eta_{3,4}$ and $\omega_{3,4}$. These parameters, as well as the moments $a_n$, are known to renormalise multiplicatively to leading log accuracy~\cite{Ball:2005ei},
\begin{equation}
c(\mu^2)=c(\mu_0^2)\left(\frac{\alpha_s(\mu^2)}{\alpha_s(\mu_0^2)}\right)^{\gamma_c/\beta_0},
\end{equation}
where $\mu_0$ is the initial scale at which the parameter was calculated and $\gamma_c$ are the one-loop anomalous dimensions defined in Tab.~\ref{tab:AnomDims} for $c=a_n$, $\eta_{3,4}$ or $\omega_{3,4}$.

\begin{table}
\hspace{-.5 cm}
 \begin{tabular}{c|c|c|c|c}
\hline\hline
\T \Bo$\gamma_{a^n}$ & $\gamma_{\eta_3}$ & $\gamma_{\omega_3}$ & $\gamma_{\eta_4}$ & $\gamma_{\omega_4}$\\
\hline
\rule[3.4ex]{0pt}{0pt}
$4 C_F \left(\psi(n+2)+\gamma_E-\frac{3}{4}-\frac{1}{2(n+1)(n+2)}\right)$ & $\frac{16}{3} C_F+C_A$ & $-\frac{25}{6}C_F+\frac{7}{3} C_A$ & $\frac{8}{3} C_F$ & $ -\frac{8}{3}C_F+\frac{10}{3} C_A$
\rule[-2.2ex]{0pt}{0pt}\\
\hline\hline
 \end{tabular}
\caption{\label{tab:AnomDims} One-loop anomalous dimensions of the parameters $a_n$, $\eta_{3,4}$ and $\omega_{3,4}$ describing the DAs~\cite{BZ04,Ball:1998je}.}
\end{table}

Coming back to the renormalisation of our NNLO result, the UV structure of the asymptotic DA can be factorised into the function $Z_\phi(u,v)$~\cite{Melic:2001wb}. This can be related to $V(u,v)$, the evolution kernel governing the renormalisation group (RG) running of the asymptotic DA, via
\begin{equation}
\hspace{-.15cm}
 V(u,v)=-\frac{1}{Z_\phi(u,v)}\left(\mu^2\frac{\partial}{\partial \mu^2} Z_\phi(u,v)\right)\,.
\end{equation}
$V(u,v)$ is defined in Refs.~\cite{Katz:1984gf,Mikhailov:1984ii}, where it was first calculated to two-loop accuracy, and is given to $\mathcal{O}(\alpha_s^2 N_f)$ by, 
\begin{equation}\label{eq:VN}
V(u,v)=\frac{\alpha_s}{2 \pi}\,V_0(u,v)+\left(\frac{\alpha_s}{2 \pi}\right)^2\frac{1}{2} N_f\,C_F\,V_N(u,v)+\ldots\,.
\end{equation}
Explicit expressions for $V_0(u,v)$ and $V_N(u,v)$ can be found in Ref.~\cite{Mikhailov:1984ii}, and the ellipsis indicates other $\mathcal{O}(\alpha_s^2)$ and higher order terms.
$Z_\phi^{(2)}(u,v)$, i.e. the $\mathcal{O}(\alpha_s^2 N_f)$ contribution to $Z_\phi(u,v)$, can then be reconstructed from the evolution kernel, and expressed in terms of $V_0(u,v)$ and $V_N(u,v)$,
\begin{equation}
Z_\phi(u,v)=\delta(u,v)+\frac{\alpha_s}{4\pi}\,\frac{1}{\epsilon}\,2V_0(u,v)+\left(\frac{\alpha_s}{4\pi}\right)^2\frac{1 }{\epsilon^2}N_fC_F \left(\frac{1}{2}V_0(u,v)+\epsilon\,V_N(u,v)\right)+\dots\,.
\end{equation}
On convolution with the tree-level hard scattering kernel $\mathcal{T}_\mu^{(2,0)}(u,\mu^2)$, i.e. the leading contribution to $\mathcal{T}_\mu^{(2)}(u,\mu^2)$ in Eq.(\ref{eq:PiT2}), the divergence up to $\mathcal{O}(\alpha_s^2 N_f)$ takes the form
\begin{equation}
\Pi_\mu^{(2),\phi_{UV}}=\int du\,\int dv\,\frac{1}{\epsilon}C_FV_N(u,v) \mathcal{T}_\mu^{(2,0)}(u,\mu^2)\phi(v,\mu^2).
\end{equation}
Note that the terms in $V_0(u,v)$ are symmetric in $u,\,v$, and therefore vanish since we use the asymptotic DA. The UV divergence of the DA cancels the IR divergence of the hard scattering kernel exactly at $\mathcal{O}(\alpha_s^2 N_f)$, i.e. $\Pi_\mu^{(2),\phi_{UV}}=-\Pi_\mu^{(2),T_\mathrm{IR}}$. Therefore the IR divergences associated with the hard-scattering kernel can be absorbed into the DA, as discussed
in detail in Ref.~\cite{Braaten:1982yp} for the case of the pion transition form-factor, leaving us with a result for $\Pi_\mu^{(2),\mathrm{ren.}}$ which is completely finite.
Convoluting this renormalised hard-scattering kernel with the asymptotic DA results in an expression including terms involving $L_4$ and generalised Nielsen polylogarithms. Since we calculate the hard scattering kernel to 
$\mathcal{O}(\alpha_s^2 N_f)$, we should take the scale dependence of the twist-2 DA to the same order, which involves adding the term $\displaystyle  2 C_F V_N(u,v) \ln(\mu^2/\mu_0^2) \Pi_\mu^{(0)}$ to the result for $\Pi_\mu^{(2)}$.

\section{Results}\label{sec:6}
Before coming to our numerical analysis, we must first extract the spectral density from the correlation function $\Pi_\mu$, and obtain the  $\mathcal{O}(\alpha_s^2 \beta_0)$ QCD sum rules result for the $B$ meson decay constant $f_B$.

\subsection{Spectral density}
As in Eq.~(\ref{eq:correlator2}), we define $\Pi^{\mathrm{T2}}_+$ in terms of $\Pi^{\mathrm{T2}}_\mu$ via
\begin{align}
\Pi^{\mathrm{T2}}_\mu=(p_B+p)_\mu \Pi^{\mathrm{T2}}_+(p_B^2,q^2)+(p_B-p)_\mu \Pi^{\mathrm{T2}}_-(p_B^2,q^2).
\end{align}
One can then extract the relevant spectral density by taking the imaginary part of the calculated correlator,
\begin{equation}
\rho_{\mathrm{T2}}=\frac{1}{\pi}\mathrm{Im} \Pi^{\mathrm{T2}}_+.
\end{equation}
An expression for the NNLO correction to $\rho_{T2}$ is given explicitly in the appendix. As we will employ the pole mass for $m_b$ in our numerical analysis, we have rewritten the $\overline{\rm MS}$ mass in terms of the pole mass. At $\mathcal{O}(\alpha_S^2N_f)$, this involved adding the term
\begin{equation}
\Delta\rho_{T2}^{(2)}=-C_f f_\pi \frac{m_b^3}{s^3} (3 m^2-2 s)\left(\frac{1}{2}(71+8 \pi^2)+26 \log\frac{\mu^2}{m^2}+6 \log^2\frac{\mu^2}{m^2}\right)\label{eq:NNLOmasscorr}
\end{equation}
to $\rho_{T2}^{(2)}$. Finally, in order to obtain the $\mathcal{O}(\alpha_S^2\beta_0)$ result, $N_f$ in $\rho_{T2}$ should be replaced by $-3/2\beta_0$. Including the contributions at twist-3 to one-loop accuracy and twist-4 to leading order accuracy,
\begin{equation}
\rho_{\Pi_+}(s,0) = \lim_{q^2\to0}(\rho_{\mathrm{T2}} +\rho_{\mathrm{T3}} + \rho_{\sigma}+\rho_{\mathrm{p}} + \rho^{2p}_{\mathrm{T4}}+\rho^{3p}_{\mathrm{T2}}),
\end{equation}
where $\rho_{\mathrm{T3}}$, $\rho_{\sigma}$ and $\rho_{\mathrm{p}}$ are contributions at twist-3 and $\rho^{2(3)p}_{\mathrm{T4}}$ are contributions at twist-4 as defined in Ref.~\cite{BZ04}. An additional twist-4 term, $\mathrm{T}4_c$, cannot be expressed via a dispersion relation so must be included separately. Therefore, on taking the Borel transformation of $\Pi_+$, we have
\begin{equation}
 \hat{B} \Pi_+=\int_{m_b^2}^\infty ds\,\rho_{\Pi_+}(s,0)e^{-s/M^2}+\mathrm{T}4_c^{(0)},
\end{equation}
where we have defined $\mathrm{T}4_c^{(0)}$ via
\begin{equation}
\mathrm{T}4_c^{(0)}=\lim_{q^2\to 0}T4_c.
\end{equation}

\subsection{Decay constant $f_B$}
Expressing the sum rule as
\begin{equation}
\label{eq:finalSR}
f_+(0)=\frac{1}{m_B^2 f_B}\left( \int_{m_b^2}^{s_0}ds\,\rho_{\Pi_+}(s,0) e^{(m_B^2-s)/M^2} +\mathrm{T}4_c\,e^{m_B^2/M^2}\right),
\end{equation}
we see that a numerical result for $f_+(0)$ requires the decay constant $f_B$ as input. For consistency we use the QCD sum rules result also calculated to $\mathcal{O}(\alpha_s^2 \beta_0)$. Although the full $\mathcal{O}(\alpha_s^2)$ corrections are sizeable~\cite{Jamin:2001fw,Penin:2001ux}, this is thought to be due to the effect of the classical Coulomb interaction~\cite{Bagan:1991sg}, such that the perturbative expansion is under control. Moreover, the same coulombic corrections would also affect the correlator for $f_+(0)f_B$. This implies that by employing the sum rules result for $f_B$ there should be a cancellation between these radiative corrections, as well as between the dependence on input parameters such as $m_b$ and $\mu$, in $f_+(0)f_B$ and $f_B$. The QCD sum rules result for $f_B$ takes the form
\begin{equation}
f_B=\frac{1}{m_B^2}\left(\int_{m_b^2}^{s_0}ds\,\rho_{\rm pert}(s) e^{(m_B^2-s)/M^2}+C_{\bar{q}q}\langle\bar{q}q\rangle+C_{\bar{q}Gq}\langle\bar{q}\sigma_g Gq\rangle\right)^{\frac{1}{2}},
\end{equation} 
where $C_{\bar{q}q}$ and $C_{\bar{q}Gq}$ are Wilson coefficients for the operator product expansion (OPE) in terms of the quark and mixed condensates respectively~\cite{Aliev:1983ra,Bagan:1991sg}. The spectral density for the perturbative contribution $\rho_{\rm pert}(s)$ can be expanded in $\alpha_s$,
\begin{equation} 
\rho_{\rm pert}(s)=\rho^{(0)}_{\rm pert}(s)+\frac{\alpha_s}{4\pi}\rho^{(1)}_{\rm pert}(s)+\left(\frac{\alpha_s}{4\pi}\right)^2 N_f \rho^{(2)}_{\rm pert}(s)\ldots ,
\end{equation}
where the tree level contribution takes the simple form 
\begin{equation}
\rho^{(0)}_{\rm pert}(s)=\frac{N_c}{8 \pi^2}m_b^2\, s\left(1-\frac{m_b^2}{s}\right)^2.
\end{equation}
The $\mathcal{O}(\alpha_s)$ result $\rho^{(1)}_{\rm pert}(s)$ was obtained from Ref.~\cite{Dominguez:1991eh}. The $\mathcal{O}(\alpha_s^2)$ corrections to $\rho_{\rm pert}(s)$, in the case that the light quark is massless, were calculated using Pad\'{e} approximations and conformal mapping and used to obtain semi-numerical results~\cite{Chetyrkin:2001je,Chetyrkin:1997mb}, as an analytical calculation of all diagrams was not feasible. We can express $\rho^{(2)}_{\rm pert}(s)$ in terms of the quantity $R^{(2),s}_{FL}(s)$, kindly provided by the authors of Ref.~\cite{Chetyrkin:2001je} in publically available code, via
\begin{equation}
\rho^{(2)}_{\rm pert}(s)=C_F\,m_b^2\,s\,R^{(2),s}_{FL}(s).
\end{equation}
To obtain the $\mathcal{O}(\alpha_s^2 \beta_0)$ result, $N_f$ in $\rho_{\rm pert}(s)$ should be replaced by $-3/2\,\beta_0$. The result for $R^{(2),s}_{FL}(s)$ is given at the scale $m_b$, and the pole mass is used for the $b$ quark. We must therefore include the $\mathcal{O}(\alpha_s^2 \beta_0)$ corrections which arise on rescaling $\alpha_s$ from $m_b$ to the factorisation scale $\mu$, which take the form
\begin{equation}
\Delta\rho^{(2)}_{\rm pert}(s)=C_F\,\mathrm{ln}\frac{m_b}{\mu}\, \rho^{(1)}_{\rm pert}(s).
\end{equation}

\subsection{Numerical analysis}
\begin{table}[tb]
\centering
 \begin{tabular}{c|c|c||c|c|c}
\hline\hline
\T \Bo
Parameter & Value & Ref.& Parameter & Value & Ref. \\
\hline\hline
\T $m_\pi$ & 139.6 MeV &\cite{Amsler:2008zzb}& $f_\pi$ & 130.4 MeV&\cite{Amsler:2008zzb}\\
$m_B$ & 5.28 GeV &\cite{Amsler:2008zzb} &$\alpha_s(M_Z)$ & 0.118 &\cite{Amsler:2008zzb}\\
$\eta_3$ & 0.015 &\cite{Ball:1998je} & $\omega_3$ & -3&\cite{Ball:1998je} \\
 $\eta_4$ & 10 &\cite{Ball:1998je}&$\omega_4$ & 0.2&\cite{Ball:1998je}\\ 
\Bo $\langle \bar{q}q\rangle$&$(-0.246^{+0.028}_{-0.019})^3 \,\mathrm{GeV}^3\,$ &\cite{Duplancic:2008zz} &$\langle \bar{q}\sigma g G q\rangle$&$(0.8\pm 0.2)\, \langle \bar{q}q\rangle$&\cite{Ovchinnikov:1988gk,Ioffe:2005ym}\\
\hline\hline
\end{tabular}
\caption{Summary of values of parameters used in the numerical analysis. Note the quark condensate is given at the scale 1 GeV.}\label{tab:Numerics}
\end{table}

From Eq.~(\ref{eq:DAexp}) it is clear that making numerical predictions for the twist-2 pion DA comes down to determining the Gegenbauer moments. This is only possible via non-perturbative methods e.g. QCD sum rules~\cite{Chernyak:1983ej,Braun:1988qv,Mikhailov:1991pt} or Lattice QCD~\cite{DelDebbio:2002mq,Braun:2006dg,Boyle:2008nj}. Recently, the UKQCD and RBC collaborations computed $a_2(2\,\mathrm{GeV})$, using $N_f=2+1$ domain-wall fermions~\cite{Arthur:2010xf}. By combining results for  $a_2(\mu)$ with experimental constraints, i.e. measurements of the $\gamma \gamma^* \pi$ form factor at CLEO~\cite{Gronberg:1997fj} and CELLO~\cite{Behrend:1991tg}, an estimate for $a_4(\mu)$ can be obtained~\cite{Stefanis:2008zi}. However, as this is a LCSR calculation, we accordingly adopt $a_{2,4}(1\,\mathrm{GeV})$ from Ref.~\cite{Khodjamirian:2011ub} where the LCSR result for the pion electro-magnetic form factor~\cite{Bijnens:2002mg} is fitted to experimental data~\cite{Huber:2008id}. The extracted values, $a_{2}(1\,\mathrm{GeV})=0.17\pm0.08$ and $a_{4}(1\,\mathrm{GeV})=0.06\pm0.10$, where the errors reflect both experimental and theoretical uncertainties, are consistent with other sum rules and Lattice QCD predictions. The parameters describing twist-3 and 4 DAs, namely $\eta_3$, $\omega_3$, $\eta_4$ and $\omega_4$, introduced in Sec.~\ref{sec:LCDAs}, were first calculated in QCD sum rules~\cite{Braun:1988qv} using non-local operator product expansion and conformal expansion. We use the updated results calculated in Ref.~\cite{Ball:1998je}, as summarised in Tab.~\ref{tab:Numerics}. The error on these parameters is taken to be $50\%$. The condensates are also required as input; we use $\langle \bar{q}q\rangle$ and $\langle \bar{q}\sigma gGq\rangle$ as given in Tab.~\ref{tab:Numerics}, neglecting the gluon condensate as its contribution is comparably small.

Our main numerical analysis is performed using the pole mass $m_b$ as input, which we calculate to $\mathcal{O}(\alpha_s^2 \beta_0)$ from the running quark mass. This improves the scale dependence of the final result, and avoids any ambiguity in the definition of the lower limit of the integral in Eq.~(\ref{eq:finalSR}). The RG improved $b$ quark mass, in the potential subtraction scheme (see Ref.~\cite{Beneke:1998rk}) was calculated at NNLO from sum rules in Ref.~\cite{Pineda:2006gx} to be  $m_b^{\mathrm{PS}}(2\,\mathrm{GeV})=4.52\pm0.06\,\mathrm{GeV}$, as in Tab. \ref{tab:Numerics}. This results in a pole mass of $4.8$ GeV at $\mathcal{O}(\alpha_s^2 \beta_0)$ (and  at $\mathcal{O}(\alpha_s^2)$), and in order not to underestimate the uncertainty on the pole mass we conservatively adopt $m_b=4.8\pm 0.1$ GeV.

\begin{figure}[t]
\begin{center}
 \includegraphics[width=0.55\textwidth]{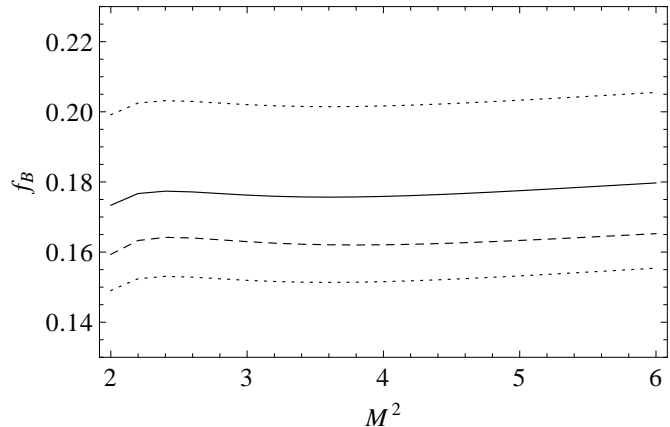}
\end{center}
\caption{$f_B(0)$ at $\mathcal{O}(\alpha_s^2\beta_0)$ as a function of the Borel parameter $M^2$, for central values of input parameters (solid) with uncertainties (dotted) calculated as described in the text for $f_+(0)$. This is compared to the $\mathcal{O}(\alpha_s)$ result calculated using $s_0=34.2 \,\mathrm{GeV}^2$ (dashed).}\label{fig:2}
\end{figure}

The LCSR approach requires a careful choice of numerical values for the continuum limit $s_0$ and the Borel parameter $M^2$. We treat the sum rules for $f_B f_+(0)$ and $f_B$ separately, obtaining independent values of $s_0$ and $M^2$ for both. These should be chosen such that the following conditions are met:
\begin{itemize}
\item  the sum rule exhibits little dependence on, but a clear extremum as a function of these parameters;
\item  the corresponding sum rule for $m_B$, which can be obtained by differentiating the sum rule for $f_B$ or $f_+(0)f_B$ by $1/M^2$, is fulfilled to $0.1\%$, as in Ref.~\cite{BZ04};
\item the continuum contribution is under control,  i.e. we impose that the integral of the spectral density between $s_0$ and $\infty$ should be approximately 25-30\% of the $B$ contribution, between $m_b^2$ and $s_0$, for $f_+(0)f_B$, and 50\% for $f_B$;
\item as far as possible, the contributions of higher orders in perturbation theory and twists should be suppressed.
\end{itemize}
Note that we rescale the Borel parameter by $\langle u \rangle^{-1}$ as defined in  Ref.~\cite{BZ04}, as the effective Borel parameter in the tree-level sum rule is $u M_{\mathrm{LC}}^2$ rather than $M_{\mathrm{LC}}^2$  corresponding to $M^2$ in Eq.~(\ref{eq:finalSR}). In our numerical analysis we find that $s_0=34.2\,\mathrm{GeV}^2$ and $M^2=3.6\,\mathrm{GeV}^2$ for $f_B$, and $s_0=34.3 \,\mathrm{GeV}^2$ and $M^2=8.1\,\mathrm{GeV}^2$ for $f_+(0)f_B$, meet the above requirements. The factorisation or renormalisation scale $\mu$ is chosen to be the typical virtuality of the $b$ quark,  $\sqrt{m_B^2-m_b^2}$, as this has previously been found to be an optimal scale~\cite{BBB98,Khodjamirian:2000ds,BZ04}. In Fig.~\ref{fig:2} we show $f_B$ as a function of $M^2$ and compare this to the corresponding result at $\mathcal{O}(\alpha_s)$.

We find that the dominant uncertainties on $f_+(0)$ arise due to varying the following:
\begin{itemize}
\item the condensates as indicated in Tab.~\ref{tab:Numerics};
\item the twist-3 parameter $\eta_3$ by $\pm 50\%$;
\item the $b$ quark mass by $\pm0.1$ GeV;
\item the continuum threshold $s_0$ by $\pm 0.5\,\mathrm{GeV}^2$ and the Borel parameter $M_2$ by $\pm 1.2\,\mathrm{GeV}^2$ for both $f_+(0) f_B$ and $f_B$;
\item the factorisation scale in the range $\mu^2\pm 2\,\mathrm{GeV}^2$.
\end{itemize}
\begin{figure}
 \centering
  \includegraphics[width=0.55\textwidth]{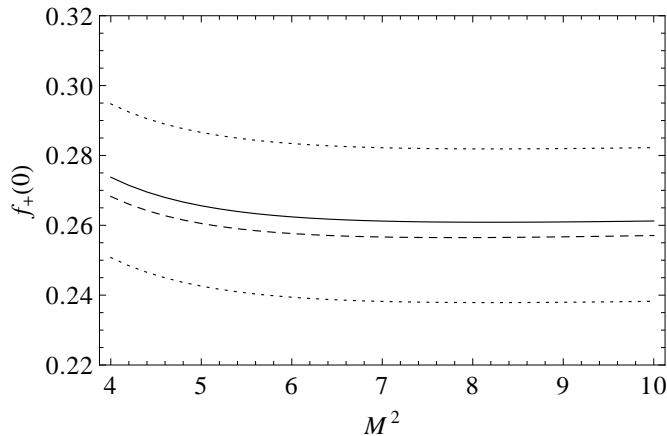}
 \caption{$f_+(0)$ at $\mathcal{O}(\alpha_s^2\beta_0)$ for central values of input parameters (solid) with uncertainties (dotted), compared to the $\mathcal{O}(\alpha_s)$ result calculated using $s_0=34.3 \mathrm{GeV}^2$ (dashed), as a function of the Borel parameter $M^2$.}\label{fig:3}
\end{figure}
The uncertainties arising from each of the above are calculated independently and added in quadrature, and we obtain $f_+(0)=0.261^{+0.020}_{-0.023}$. The uncertainties are less than $9\%$, and could be further reduced by better determining the condensates and the twist-3 parameters via, for example, Lattice QCD. Comparing our result for $f_+(0)$ to the $\mathcal{O}(\alpha_s)$ result in Fig.~\ref{fig:3} shows that, despite the $\sim 9\%$ $\mathcal{O}(\alpha_s^2\beta_0)$ corrections to $f_B$ mentioned earlier, there is little change in $f_+(0)$ $\sim 2\%$. This observation indicates the reliability of the light-cone sum rule approach to the calculation of form factors, as it seems that the results are stable with respect to higher order corrections. This could further be taken as confirmation that the QCD sum rules result for $f_B$ should indeed be used in preference to the Lattice QCD result in LCSR calculations of the form factors.

In Ref.~\cite{Duplancic:2008zz,Khodjamirian:2011ub}, $f_+(0)$ was calculated using the $b$ quark mass in the $\overline{\mathrm{MS}}$ scheme. Here it was argued that this is a natural scheme for the calculation of scattering amplitudes involving a virtual $b$ quark at large space-like momentum scales $\sim m_b$. As there are arguments in favour of both schemes, we also calculate our result using the $\overline{\mathrm{MS}}$ mass for the $b$ quark. This involves replacing the pole mass by the $\overline{\mathrm{MS}}$ mass at the scale $\mu$, adding NLO corrections found in the appendix of Ref.~\cite{Duplancic:2008zz} for both twist-2 and 3 scattering kernels. At NNLO translating back to the $\overline{\mathrm{MS}}$ scheme for the $b$ mass means removing the correction given in Eq.~(\ref{eq:NNLOmasscorr}).
As for $f_B$, we take the expressions given in Ref.~\cite{Jamin:2001fw} up to $\mathcal{O}(\alpha_s^2\beta_0)$. For the value of the mass, we use $m_b(m_b)=4.19^{+0.18}_{-0.06}$~\cite{PDG}. Note that, as in Ref.~\cite{Jamin:2001fw}, we use the pole mass for the continuum cut-off, although using the running mass here instead would change our result negligibly. Imposing the same requirements as for the pole-mass scheme, we find $s_0=35.3\,\mathrm{GeV}^2$ and $M^2=3.7\,\mathrm{GeV}^2$ for $f_B$, and $s_0=35.7 \,\mathrm{GeV}^2$ and $M^2=7.8\,\mathrm{GeV}^2$ for $f_+(0)f_B$, and obtain $f_+(0)=0.252^{+0.019}_{-0.028}$. This is $\sim 3\%$ below the result in the pole-mass scheme.

\subsection{Determination of $|V_{ub}|$}
As mentioned in the introduction, it is possible to predict $|V_{ub}|$ using the experimental determination of $f_+(0)|V_{ub}|$ and $f_+(0)$ from LCSR. In Ref.~\cite{Ball:2006jz} $f_+(0)|V_{ub}|$ was first obtained by fitting various form-factor shape parameterisations to BaBar data~\cite{Aubert:2006px}. It was observed that the results for $f_+(0)|V_{ub}|$ were independent of the parameterisation method chosen. Recently BaBar and Belle quote results for $f_+(0)|V_{ub}|$, extracted by fitting binned data to a Boyd-Grinstein-Lebed~\cite{Boyd:1994tt} or Becirevic-Kaidalov~\cite{Becirevic:1999kt} parameterisation respectively, as summarised in Tab.~\ref{tab:Exp} along with the corresponding value of $|V_{ub}|$. Where necessary, the statistical and systematic uncertainties are added in quadrature.
\begin{table}
\centering
 \begin{tabular}{c|c|c|c|c}
\hline
\hline
 \T \Bo Exp. & No. of bins & Ref. &$f_+(0)|V_{ub}|$ & $|V_{ub}|$\\
\hline
\hline
\T\Bo  BaBar &  6 & \cite{Babar:2010uj} &$(1.08\pm0.06)10^{-3}$&$(4.13^{+0.36}_{-0.32}|_{\rm th.}\pm 0.23|_{\rm exp.})10^{-3}$\\
\T\Bo  BaBar &  12& \cite{delAmoSanchez:2010zd}&$(8.6\pm 0.3_{\rm stat}\pm 0.3_{\rm syst})10^{-4}$&$(3.29^{+0.29}_{-0.26}|_{\rm th.}\pm 0.16|_{\rm exp.})10^{-3}$\\
\T\Bo  Belle &  13 & \cite{Ha:2010rf}&$(9.24\pm 0.18_{\rm stat}\pm 0.21_{\rm syst})10^{-4}$&$(3.54^{+0.31}_{-0.28}|_{\rm th.}\pm 0.11|_{\rm exp.})10^{-3}$\\
\hline\hline
 \end{tabular}
\caption{Predictions of $|V_{ub}|$ using $f_+(0)|V_{ub}|$ from analyses in 2010 of $B\to\pi l \nu$ data.}\label{tab:Exp}
\end{table}
We find that although there is a slight tension between Refs~\cite{Babar:2010uj} and \cite{delAmoSanchez:2010zd}, these predictions are on the whole in keeping with the CKMFitter result~\cite{Charles:2011va}, $|V_{ub}|=(3.501^{+0.196}_{-0.087})10^{-3}$ and the UTFit result~\cite{Bona:2010}, $|V_{ub}|=(3.64\pm 0.11)10^{-3}$. They are also in good agreement with the most recent value obtained from LCSR in Ref.~\cite{Khodjamirian:2011ub}, $|V_{ub}|=(3.50^{+0.38}_{-0.33}|_{\rm th.}\pm 0.11|_{\rm exp.})10^{-3}$. We look forward to the results from SuperB and Super-KEKB which should enable further improvements on the precision of the exclusive determination of $|V_{ub}|$. 

\section{Summary}\label{sec:7}
We have calculated the $\mathcal{O}(\alpha_s^2 \beta_0)$ corrections to $f_+(0)$ at leading-twist in QCD sum rules on the light-cone, and  performed a comprehensive numerical analysis of the result, including NLO twist-3 and LO twist-4 contributions, leading to a new determination of $V_{ub}$. We have found that in spite of $\sim 9\%$ positive NNLO corrections to the QCD sum rules result for $f_B$ seen in Fig.~\ref{fig:2}, the LCSR prediction for $f_+(0)$ is stable, increasing by $\sim 2\%$ to $f_+(0)=0.261^{+0.020}_{-0.023}$, as shown in Fig.~\ref{fig:3}. This increases our confidence in the stability of LCSR calculations for form factors with respect to this source of radiative corrections, and provides further indication that in the calculation of the form factors in LCSR, $f_B$ should be taken from sum rules rather than Lattice QCD. We find that on inclusion of our NNLO correction, the scale dependence is reduced, and the main sources of theoretical uncertainty are due to $a_2$ and $m_b$. The total uncertainty of $\sim 9\%$ could be reduced  in the future by the determination of the condensates and twist-3 parameters on the Lattice. We also evaluate $f_+(0)$ using the $\overline{\rm MS}$ mass for the $b$ quark and find $f_+(0)=0.252^{+0.019}_{-0.028}$, in agreement with the result obtained using the pole mass. Finally predictions for $|V_{ub}|$ were obtained in the range $(3.29-4.13)\cdot 10^{-3}$, making use of $f_+(0)|V_{ub}|$ from BaBar and Belle, in Tab.~\ref{tab:Exp}. We stress that our approach to $f_+(0)$ in LCSR is complementary to Lattice QCD calculations of $f_+(q^2)$ as the latter technique is more applicable to the region of large $q^2$.  Therefore the determination of $|V_{ub}|$ by fitting both our result along with Lattice predictions to the combined experimental results~\cite{Bourrely:2008za,Flynn:2007qd} would also be of great interest.

\section*{Acknowledgments}
The author is very grateful to Patricia Ball for her guidance throughout this project, to Thorsten Feldmann for a careful reading of the draft and helpful suggestions, and to Roman Zwicky, Adrian Signer and Vladimir Braun for enlightening discussions. The early part of this work was supported by a STFC studentship, and later by the DFG grant SFB 676, ``Particles, Strings, and the Early Universe''.

\appendix

\section*{Appendix}

In analogy to Eq.~(\ref{eq:PiT2}), we can perturbatively expand the twist-2 spectral density,
\begin{equation}
\rho_{\mathrm{T2}}=\rho_{\mathrm{T2}}^{(0)}+\frac{\alpha_s}{4 \pi}\,\rho_{\mathrm{T2}}^{(1)}+\left(\frac{\alpha_s}{4 \pi}\right)^2 N_f\,\rho_{\mathrm{T2}}^{(2)}\dots\,.
\end{equation}
Our NNLO correction $\rho_{\mathrm{T2}}^{(2)}$ then takes the form,
\begin{align}
\rho_{\mathrm{T2}}^{(2)}=&f_\pi C_F\bigg\{
\frac{5m_b^3(m_b^2-s)}{3 s^3}\log^3\left(1-\frac{m_b^2}{s}\right)\nonumber \\
&+\left(\frac{9m_b^3(m_b^2-s)}{s^3}\log\left(\frac{s}{m_b^2}\right)-\frac{m_b(m_b^2-s)(20 m_b^4-42 s m_b^2+7 s^2)}{2 s^4}\right)\log^2\left(1-\frac{m_b^2}{s}\right)\nonumber \\
&+\bigg(\frac{9m_b^3(m_b^2-s)}{s^3}\log^2\left(\frac{s}{m_b^2}\right)-\frac{2m_b(36 m_b^6-47 s m_b^4+35 s^2 m_b^2-6 s^3)}{3 s^4}\log\left(\frac{s}{m_b^2}\right)\nonumber \\
&+\frac{m_b(m_b^2-s)(32 m_b^4-10\pi^2s m_b^2-428 s m_b^2+79 s^2)}{6 s^4}\nonumber\\
&-\frac{14m_b^3(m_b^2-s)}{s^3}\mathrm{Li}_2\left(\frac{m_b^2}{s}\right)\bigg)\log\left(1-\frac{m_b^2}{s}\right)\nonumber\\
&+\frac{m_b^3(m_b^2-s)}{3s^3}\log^3\left(\frac{s}{m_b^2}\right)-\frac{m_b^3(48m_b^4-134 s m_b^2+107s^2)}{6 s^4}\log^2\left(\frac{s}{m_b^2}\right)\nonumber \\
&-\frac{6 m_b^3(m_b^2-s)}{s^3}\log^2\left(\frac{m_b^2}{s}-1\right)+\log\left(\frac{\mu}{\mu_0}\right)\bigg(-\frac{2 m_b^3(m_b^2-s)}{s^3}\log^2\left(1-\frac{m_b^2}{s}\right)\nonumber \\
&-\left(\frac{4m_b^3(m_b^2-s)}{s^3}\log\left(\frac{s}{m_b^2}\right)+\frac{2 m_b(m_b^2-s)}{s^2}\right)\log\left(1-\frac{m_b^2}{s}\right)\nonumber \\
&-\frac{2 m_b^3(m_b^2-s)}{s^3}\log^2\left(\frac{s}{m_b^2}\right)-\frac{2m_b^3}{s^2}\log\left(\frac{s}{m_b^2}\right)+\frac{2 m_b^3(-15+\pi^2)(m_b^2-s)}{3 s^3}\bigg)\nonumber \\
&+\frac{2 m_b(m_b^2-s)(8 m_b^4-8 s m_b^2+s^2)}{s^4}\log\left(\frac{s}{m_b^2}-1\right)\nonumber\\
&+\frac{2m_b(6 m_b^6+22sm_b^4-19s^2m_b^2-s^3)}{s^4}\mathrm{Li}_2\left(\frac{m_b^2}{s}\right)\nonumber
\end{align}

\begin{align}
&-\log\left(\frac{s}{m_b^2}\right)\bigg(\frac{m_b(-32 m_b^6-8\pi^2sm_b^4+484sm_b^4+8\pi^2s^2m_b^2-239s^2m_b^2+24s^3)}{6s^4}\nonumber\\
&+\frac{8 m_b^3(m_b^2-s)}{s^3}\log\left(\frac{s}{m_b^2}-1\right)+\frac{18 m_b^3(m_b^2-s)}{s^3}\mathrm{Li}_2\left(\frac{m_b^2}{s}\right)\bigg)\nonumber \\
&-\log\left(\frac{\mu}{m_b^2}\right)\bigg(\frac{2m_b^3(m_b^2-s)}{s^3}\log^2 \left(1-\frac{m_b^2}{s} \right)\nonumber\\
&-\left(\frac{2m_b(m_b^2-s)(4m_b^4-10sm_b^2+s^2)}{s^4}-\frac{8m_b^3(m_b^2-s)}{s^3}\log\left(\frac{s}{m_b^2}\right)\right)\log\left(1-\frac{m_b^2}{s}\right)\nonumber\\
&+\frac{2m_b(m_b^2-s)(\pi^2m_b^2-27m_b^2+9s)}{3s^3}-\frac{2m_b^3(4m_b^4-14sm_b^2+9s^2)}{s^4}\log\left(\frac{s}{m_b^2}\right)\nonumber \\
&-\frac{12 m_b^3(m_b^2-s)}{s^3}\mathrm{Li}_2\left(\frac{m_b^2}{s}\right)\bigg)\nonumber \\
&+6\frac{ m_b^3(m_b^2-s)}{s^3}\mathrm{Li}_3\left(1-\frac{m_b^2}{s}\right)-\frac{10m_b^3(m_b^2-s)}{s^3}\mathrm{Li}_3\left(\frac{m_b^2}{s}\right)\nonumber \\
&+\frac{m_b}{9s^4}\bigg(18\pi^2m_b^6-137\pi^2sm_b^4+363sm_b^4-216s\zeta(3)m_b^4+98\pi^2s^2m_b^2\nonumber \\
&-561s^2m_b^2+216s^2\zeta(3)m_b^2-3\pi^2s^3+198s^3\bigg) 
\bigg\},
\end{align}
where $\mu_0$ is the scale at which the DA moments are calculated.

\end{document}